\documentclass[doublecol,figures]{epl2}

\title
{Two-fluid behaviour at the origin of the resistivity peak in  
doped manganites}
\shorttitle
{2-fluid behaviour at the origin of the resistivity peak in 
manganites}

\author{D. I. Golosov\inst{1}\thanks{E-mail: \email{golosov@phys.huji.ac.il}} 
\and N. Ossi\inst{1} \and A. Frydman\inst{1} \and I. Felner\inst{2} \and
I. Nowik\inst{2} \and  M. I. Tsindlekht\inst{2} \and Y. M. Mukovskii\inst{3}}
\shortauthor{Golosov \etal}
\institute{
\inst{1} Department of Physics and the Resnick Institute, Bar-Ilan 
University, Ramat-Gan 52900, Israel \\
\inst{2} Racah Institute of Physics,
The Hebrew University, Jerusalem 91904, Israel\\
\inst{3}Moscow State Steel and Alloys
Institute, 119049 Moscow, Russia
}

\abstract{
We report a series of magnetic and transport measurements on high-quality
single crystal samples of colossal magnetoresistive manganites,
${\rm La_{0.7} Ca_{0.3} Mn O_3}$ and ${\rm Pr_{0.7} Sr_{0.3} Mn O_3}$. 
1 \% Fe
doping allows  a M\"{o}ssbauer spectroscopy study, which shows 
(i) unusual
line broadening within the ferromagnetic phase and (ii) a coexistence of 
ferro- and 
paramagnetic contributions in a region, $T_1<T<T_2$, around the 
Curie point $T_C$. In the case of 
${\rm Pr_{0.7} Sr_{0.3} Mn O_3}$, the resistivity peak 
occurs at a considerably higher temperature, $T_{MI}>T_2$. This shows that 
phase separation 
into metallic (ferromagnetic) and insulating
(paramagnetic) phases cannot be generally responsible for the 
resistivity  peak (and hence for the associated colossal magnetoresistance). 
Our results can
be understood phenomenologically within the two-fluid approach,
{which also allows for a difference between $T_C$ and $T_{MI}$. 
Our  data indeed 
imply that while  magnetic and transport properties of the manganites
are closely interrelated, the two transitions at $T_C$ and $T_{MI}$  
can be viewed as  distinct
phenomena.} 
}
\begin{document}

\maketitle

\section{Introduction}
The phenomenon of colossal magnetoresistance (CMR) in doped manganese
oxides continues to attract extensive research 
effort\cite{Tokurabook}.
While the physical mechanism underlying the CMR phenomenon remains elusive,
one can expect it to be generic for the entire family of the CMR compounds,
spanning a broad range of chemical compositions, dopant levels, and lattice
properties. This natural suggestion is corroborated by the fact that, in 
addition to
the CMR itself,
other generic unusual features of the CMR compounds have been found in 
recent years. These include the formation of a ``pseudogap'' in the carrier
density of states on increasing temperature toward $T_C$ \cite{pseudogap} 
(arguably responsible for the peak in the
resistivity, $\rho(T)$, which in turn gives rise to the CMR 
effect\footnote{Note also the optical\cite{okimoto} and transport\cite{chun} data,
suggesting that the effective carrier number decreases with $T$ increasing
toward $T_C$.}), and the 
unusual short-range magnetic correlations
in the critical region (``central peak''), observed in the 
inelastic
neutron scattering experiments\cite{centralpeak}. Although these 
findings testify to the
anomalous electronic and magnetic properties respectively, the relationship
between the two remains unclear. While it is generally understood that the 
metal-insulator
transition (at $T_{MI}$, corresponding to the resistivity peak) and the
ferro- to paramagnetic transition (at the Curie temperature $T_C$) lie close
to each other, the consensus does not go much further. Indeed, questions
relating to the mutual location of the two 
transitions\cite{Nagaevreview,yamtc} 
(whether $T_C$ actually {\it equals} $T_{MI}$), as well as to their character, 
remain largely open.
This situation is partly due to sample preparation issues, which make it
difficult to draw a  quantitative connexion between results
of different measurements, carried out on different samples.

On the theory side, the picture is similarly uncertain. Current 
ideas on the mechanism of the CMR mainly fall within two general categories, 
which 
can be referred to phenomenologically without specifying the 
microscopic models or even the relevant degrees of freedom. The
first one is the 
so-called {\it phase separation scenario}\cite{phasep,Dagotto}, 
whereby the 
metallic
(ferromagnetic) and insulating (paramagnetic) phases coexist in the 
temperature region around $T_C$. With decreasing temperature, the volume
fraction of ferromagnetic ``droplets'' grows, resulting in an eventual 
percolation and metallic behaviour at low $T$. The long-range 
ferromagnetic order is also linked to connectivity between the ferromagnetic
clusters and is established at about the same point.

The second theoretical view is that of 
the {\it two-fluid model}\cite{Rama}, based on the 
coexistence\cite{Salamon} of 
(i) localised carriers (or polarons \cite{Millis}) and
(ii) itinerant conduction electrons (present below $T_{MI}$), in a 
spatially 
homogeneous (on a submicron scale) system. While the 
high-temperature insulating properties are accounted for by the fact that
all carriers are localised, lowering the temperature  leads
to an increase of the itinerant carrier population. This in turn results in (i)
the resistivity passing through a maximum and decreasing toward lower $T$
and (ii) strengthening the ferromagnetic interaction in the system via
the double exchange mechanism\cite{degennes}, which is more effective
for the itinerant electrons than for the localised ones\footnote{Unless 
the electron is localised on a single lattice site, the
localised carriers also contribute to double exchange ferromagnetism. This
effect is however weaker than for the itinerant electrons.}.
%\cite{local}. 
{We note that the prevalent view\cite{Millis95} 
is that the double
exchange mechanism on its own cannot lead to localisation of carriers
above certain temperature, hence a microscopic model of colossal 
magnetoresistance must include additional interactions and/or additional 
degrees of freedom.}
In principle, the  ferromagnetic order can be established above or 
below the
downturn of the resistivity.
Thus, while the physics behind the magnetic transition at $T_C$ can be that
of thermal fluctuations overpowering 
the ferromagnetism, the
electronic transition at $T_{MI}$ is due to a (presumably strongly correlated) mechanism 
leading to the disappearance of itinerant carriers at higher $T$.
{One may expect this latter mechanism to become operational
only in the presence of sufficiently
strong magnetic fluctuations (as found near $T_C$), so that ultimately
the two transitions are inter-related}.

In the present letter, we report a series of magnetic and transport 
measurements. All of these were performed on  the {\it same} 
high-quality manganite single crystals. Our findings appear 
incompatible
with the percolative nature of the metal-insulator transition 
(as in the phase separation scenario) being a general property. 
They also imply that the ferro- to 
paramagnetic and electronic (metal-insulator) transitions are two distinct
phenomena with no rigid interconnection. Indeed, the metal-insulator transition
at $T_{MI} \neq T_C$ leads to a change of the Curie--Weiss temperature and
therefore to a separate feature in the {\it magnetic} properties of
the system at $T=T_{MI}$. These results can be understood in the framework
of the two-fluid approach.  

While the two compounds,  ${\rm Pr_{1-x} Sr_{x} Mn O_3}$ (PSMO) and
 ${\rm La_{1-x} Ca_{x} Mn O_3}$ (LCMO) with $0.2 \stackrel{<}{\sim} x \stackrel{<}{\sim} 0.5 $ show
metallic behaviour at low $T$ and CMR near $T_C$, their properties
differ significantly. The unconventional magnetic behaviour 
(giving rise to the central peak as observed near $T_C$
in the neutron scattering experiments\cite{centralpeak}),
first found in LCMO, is
much less pronounced in 
the 
PSMO\cite{centralpeak}. The 
underlying unusual short-range magnetic 
correlations
 should also be accessible via a local
({\it i.e.,} momentum-integrated) magnetic probe, such as 
M\"{o}ssbauer spectroscopy of iron-doped samples. Earlier 
M\"{o}ssbauer studies of the CMR 
manganites\cite{moessgeneral,ChecherskyFNT,greeks} 
indeed uncovered
an unusual coexistence
of para- and ferromagnetic contributions to the hyperfine field near 
$T_C$ ,which can be expected to be a manifestation of the same 
phenomenon. Other local magnetic probes,
such as muon scattering\cite{muon} and nuclear magnetic resonance\cite{nmr}, 
also find several different 
contributions in the region around $T_C$.

\section{Experimental}
Single crystals of pure and ${\rm ^{57}Fe}$-doped 
PSMO and LCMO (with $x=0.3$) were grown by non-crucible 
floating-zone melting with radiation heating\cite{furnace}. 
X-ray diffraction and EPMA (Electron Probe Micro Analysis)
measurements were
performed to verify that the samples are single phase crystals of the nominal
compositions. Properties of the samples are summarised in table
\ref{tab:samples}. 
As expected\cite{iron}, the 1\% substitution of Mn by enriched 
${\rm ^{57}Fe}$ 
(needed for M\"{o}ssbauer spectroscopy) 
in the two 
%PSMO-F %\cite{psmof} and LCMO-F 
samples,  PSMO-F  and LCMO-F, does not
significantly change the respective system properties\footnote{The resistivity of PSMO (fig. \ref{fig:mit}, top) is
indeed very close to that of PSMO-F; see also table \ref{tab:samples}.} .   

\begin{largetable}
\caption{Samples -- compositions and measured properties, including: Curie
and paramagnetic Curie--Weiss temperatures, $T_C$ and $T_{CW}$ (from magnetic 
measurements);
metal-insulator transition temperature, $T_{MI}$ and transport
gap $\Delta$, from the transport experiments. 
Coexistence of
ferro- and paramagnetic contributions to M\"{o}ssbauer spectra takes place at $T_1<T<T_2$.} 
\label{tab:samples}
%\begin{ruledtabular}
\begin{tabular}{llllllll}
Sample & Composition & $T_1$,K & $T_2$,K &$T_C$, K &$T_{MI}$, K &$\Delta$, eV& $T_{CW}$,K \\
PSMO   & ${\rm Pr_{0.7} Sr_{0.3} Mn O_3}$   &         &         &  218    & 245        &  0.068     &  205 \\
PSMO-F &  ${\rm Pr_{0.7} Sr_{0.3} Mn_{0.99} Fe_{0.01} O_3}$           & 180     &   230   &  223    &   244      &  0.064     &  221 \\
LCMO-F &   ${\rm La_{0.7} Ca_{0.3} Mn_{0.99} Fe_{0.01} O_3}$          &   190   &  230    &  221    &  231       &   0.078   &  222 
\end{tabular}
%\end{ruledtabular}
\end{largetable}

Magnetisation measurements were performed in a commercial (Quantum Design)
SQUID magnetometer. The in-phase component of the zero-field ac 
susceptibility (measured 
at  1065 Hz, amplitude 0.05 Oe) and
the standard four contact resistivity were measured by homemade probes inserted
into the magnetometer. Pieces of LCMO-F and PSMO-F crystals were crushed to 
powder for the M\"{o}ssbauer measurements. These were performed using a constant 
acceleration drive 
in transmission mode and a 50 mCi ${\rm ^{57}Co:Rh}$ source.
Measurements taken at 4.2 K
revealed  well-defined sextet spectra 
(fig. \ref{fig:moesspectra}, bottom), indicating that the Fe ions were located at a specific lattice
site (presumably that of Mn). Further measurements were performed between 90K 
and 300 K.   M\"{o}ssbauer spectra  were fitted to a superposition of
magnetic (with an asymmetric Gaussian distribution of hyperfine fields) and
paramagnetic (quadrupole doublet) components.
%(temperature stabilisation to 1K), 

\section{Results and discussion}
\begin{figure}
\includegraphics[bb=60 260 688 738, clip, scale=0.49]{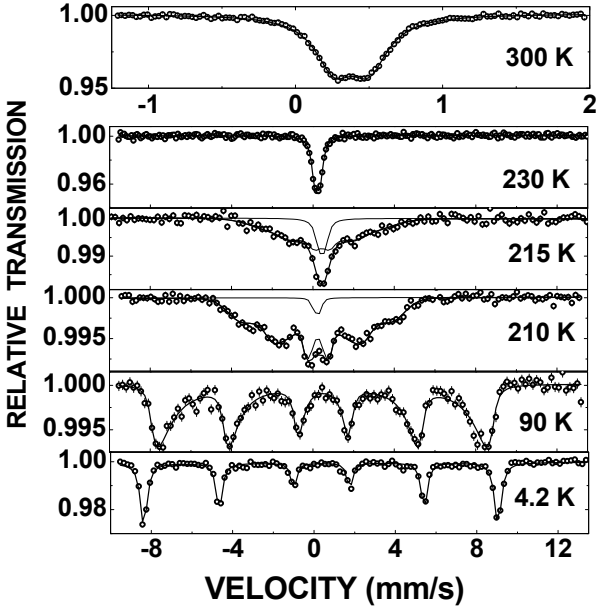}
\caption{A selection of 
M\"{o}ssbauer spectra obtained for 
PSMO-F at various temperatures. Thin solid lines represent
the simulated sub-spectra; notice the coexistence of ferro- and paramagnetic
contributions at 210K and 215 K.} 
\label{fig:moesspectra} 
\end{figure}

\begin{figure}
\includegraphics[bb=128 318 690 815, clip, scale=0.44]{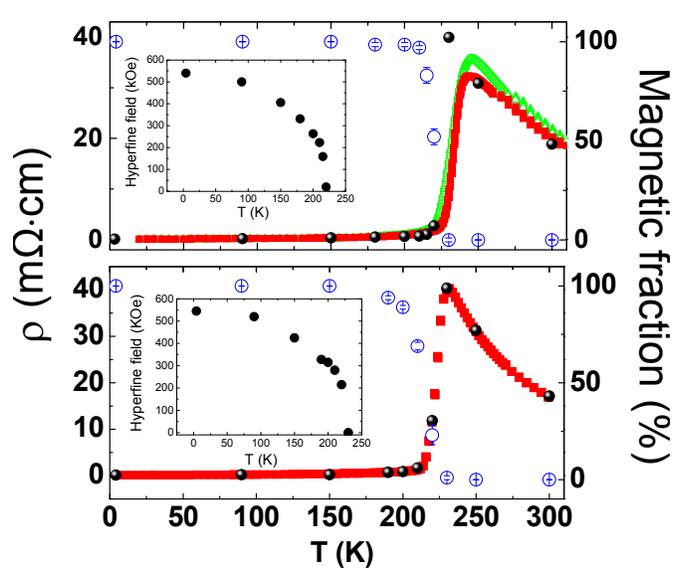}
\caption{(colour online) M\"{o}ssbauer and transport data.
Magnetically ordered fraction (open circles with error bars, 
right scale) and resistivity  $\rho(T)$ (filled squares, left scale)
for the PSMO-F (top) and LCMO-F (bottom) samples. 
Triangles (top) show $\rho(T)$ for the pure PSMO sample. Bullets correspond 
to the
effective-medium\cite{griffiths} values of $\rho(T)$ for PSMO-F and LCMO-F. 
Temperature dependences
of (most probable) magnetic hyperfine fields are plotted in the insets.}
\label{fig:mit}
\end{figure}

Our M\"{o}ssbauer results are represented in figs. \ref{fig:moesspectra} and
\ref{fig:mit}. 
Fig. \ref{fig:moesspectra} shows a sequence of M\"{o}ssbauer
spectra for the PSMO-F sample (our spectra for LCMO-F look similar to those 
shown in refs. \cite{moessgeneral,ChecherskyFNT}). While the 
spectra at 90 K and 230 K show the
magnetically ordered (sextet) and paramagnetic (doublet,
also shown on a larger scale  at 300K) 
%room temperature, fig. \ref{fig:moesspectra}, top) 
behaviour, 
respectively, a superposition of
the two contributions is clearly observed between $T_1 \approx 180 K$ and 
$T_2 \approx 230 K$
(cf. fig. \ref{fig:mit}).
Above $T_1$, both the relative intensity of the ferromagnetic
component 
(fraction of magnetically ordered sites, fig. \ref{fig:mit}, shown also 
for LCMO-F) and 
the value of the 
hyperfine field  (for both PSMO-F and LCMO-F) characterising the ferromagnetic contribution 
decrease and ultimately 
vanish at $T_2$.   
The long-range magnetic order, on the other hand, 
disappears\footnote{We defined $T_C$ as a point where the derivative 
of the magnetic susceptibility, 
${\it \partial {\rm Re} \chi /\partial T}$, is minimal.} at 
$T_C=223$, $T_1<T_C<T_2$.
%\cite{chi}

\begin{figure}
\includegraphics[bb=70 295 756 812, clip, scale=0.35]{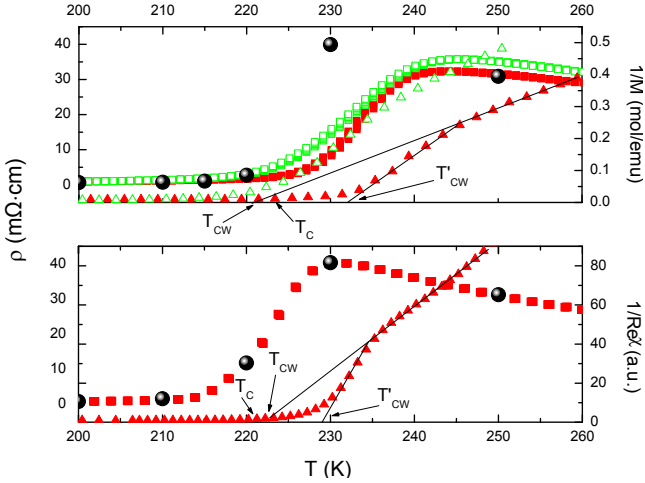}
\caption{(colour online) Filled triangles show the 
inverse low-field (10-20 Oe) magnetisation $1/M$ for PSMO-F (top) and
inverse ac susceptibility $1/{\rm Re}\chi$ for LCMO-F (bottom). Filled 
squares
represent corresponding values of $\rho(T)$, whereas bullets are the 
effective medium result for $\rho$. Open symbols (top) 
show $1/M$ and $\rho$ for PSMO; {values of $T_C$, $T_{CW}$ and  
$T^\prime_{CW}$
are shown for PSMO-F (top) and LCMO-F (bottom);} the lines are a 
guide to the eye.}
\label{fig:magnet}
\end{figure}

We find that for PSMO-F  the ferromagnetic fraction disappears more 
abruptly
(mostly at $210K <T<230K$, see fig. \ref{fig:mit}, top) 
{than for LCMO-F}, 
in agreement 
with the
more conventional critical properties found by neutron 
scattering\cite{centralpeak}. Note that the increase of resistivity $\rho$ 
with $T$ (for $T \stackrel{<}{\sim} T_{MI}$) is steeper in the case of
LCMO-F (figs.\ref{fig:mit} and \ref{fig:magnet}). Phase separation scenario
would imply the opposite trend. 
We will now consider the behaviour of $\rho(T)$
in more detail.

Our low-T resistivity data are best fitted by\cite{Schiffer}
$\rho=\rho_0 + \rho_1 T^{5/2}$.
With increasing temperature,  $\rho(T)$ passes through a maximum 
at $T=T_{MI}$, followed by an activated-type
dependence, $\rho \sim \exp{(\Delta/k_B T)}$ with $\Delta \sim 0.07 $ eV 
at higher
$T$.  In the case of PSMO-F, magnetically ordered  sites are absent already at 
$T_2=230K$, some 15 K below $T_{MI}$. This  is similar
to  earlier results for   
${\rm La_{0.8} Ca_{0.2} Mn O_3}$ 
(with $T_{MI}-T_2 \approx 7K$)\cite{ChecherskyFNT}. 
{\it Such behaviour is impossible to 
reconcile with the phase separation scenario}, which implies that the 
disappearance of metallic
ferromagnetic areas occurs {\it at or above} $T_{MI}$. Indeed, the 
effective-medium calculation\cite{griffiths,McLachlan} for resistivity
at a given value of metallic fraction suggests that $T_{MI}$
is well below the actual value (figs. \ref{fig:mit} and \ref{fig:magnet}).

In the case of LCMO-F, M\"{o}ssbauer spectra at 230 K (only slightly below 
$T_{MI}$) still indicate some presence of magnetically ordered sites 
[1($\pm 1$)\% fraction].
The resistivity data are in a perfect
agreement with the effective medium description. This is in line
with earlier results\cite{greeks} for ${\rm La_{0.67}Ca_{0.33}Mn_{0.99}
Sn_{0.01}O_3}$.
However, our results for PSMO-F (see above) prove that {\it this situation
is not generic}.
{\it The apparent 
success of phase-separation
scenario in describing LCMO-F should be viewed as incidental},
reflecting the relatively low value of $T_{MI}-T_C$ for this case (table
\ref{tab:samples}). It is plausible that the (nearly) first-order 
transition\cite{centralpeak,lynnlcmo} observed in LCMO  (with
$x\geq 0.3$) at $T_C$ circumvents a 
more generic behaviour and is indeed accompanied by phase separation, 
thereby masking the physics behind the resistivity peak. 

We note an earlier suggestion\cite{ChecherskyFNT}
that in the case of $T_2 < T_{MI}$ (realised in 
${\rm La_{0.8} Ca_{0.2} Mn O_3}$  \cite{ChecherskyFNT} and in our PSMO-F),
ferromagnetic phase (assumed metallic)
may form filaments, which facilitate metallic conductance down to a low
value of their relative volume.
While this view appears to allow for an interpretation of the results in terms 
of phase separation
scenario, we find that in the case of PSMO-F, the metallic volume fraction 
at $T>230{\rm K}$ would
have to be well below  1 \%. Such behaviour does not seem plausible both on  
energy
grounds (large surface energy term) and because one would expect such a 
structure to show a hysteretic behaviour and/or history dependence. These, 
however, were not found in our transport and magnetisation measurements.
In addition, the effective media equation\cite{griffiths,McLachlan} which is 
so successful in 
quantitatively describing
the resistivity peak in LSMO-F  (fig. \ref{fig:mit}) assumes {\it spherical} 
inclusions of minority phase,
and it is extremely unlikely that geometry of such inclusions should change
so drastically in the case of PSMO-F or ${\rm La_{0.8} Ca_{0.2} Mn O_3}$. 
We therefore conclude that both in our
PSMO-F sample
%(with $T_{MI}-T_2 \stackrel{>}{\sim} 14 {\rm K}$) 
and in  the
${\rm La_{0.8} Ca_{0.2} Mn O_3}$ ceramics of ref. \cite{ChecherskyFNT} 
%(with $T_{MI} - T_2 \approx 7{\rm K}$) 
phase separation scenario cannot account for the resistivity
peak.

An alternative interpretation is provided by the two-fluid approach, which 
allows for the itinerant electrons in the paramagnetic phase. 
On the other hand,
continued presence of  localised electrons below $T_2$ may explain
the origin of the paramagnetic contribution to the M\"{o}ssbauer spectra.
Between the sites where 
the wave function of a localised electron is  centred, 
the  ferromagnetic  double exchange coupling is weaker than 
elsewhere\footnotemark[2],
%\cite{local}, 
hence the corresponding spins order at a lower $T<T_2$. 
Even below $T_1$,
when uniform ferromagnetic order is maintained, thermal fluctuations of
these spins will be stronger, corresponding to a broad distribution
of the hyperfine field values. Indeed, we  observe an asymmetric 
line broadening in the M\"{o}ssbauer spectra 
of both LCMO-F and 
PSMO-F at ${\rm 90K}<T<T_2$ (with linewidths, for $T>150K$, reaching 
over 50\% of the 
hyperfine field value). {While inhomogeneities, which appear 
inherent in the
single crystals of doped manganites, might result in smearing 
of
the Curie transition by a few degrees K ({\it i.e.,} in a 
distribution of $T_C$ values within 
the sample, see, {\it e.g.}, ref. 
\cite{yamtc}), this cannot possibly account for such strong line
broadening well below $T_C$.}
We note that the earlier M\"{o}ssbauer 
studies\cite{moessgeneral,ChecherskyFNT,greeks}, carried out typically on
ceramic samples (whose measured properties can be affected by the inter-grain
boundaries)\footnote{We are aware of a sole exception\cite{moesssingle}, 
reporting emission M\"{o}ssbauer data for a single crystal of 
${\rm La_{0.9} Ca_{0.1} Mn O_3}$, 
which is well outside the metallic doping range.},
%\cite{moesssingle} 
suggested the coexistence of ferro- and 
paramagnetic
contributions down to lower $T$, and/or the presence of several distinct ferro-
(at low $T$) or paramagnetic (at high $T$) phases. The continuous 
hyperfine field
distribution observed in our samples (also at $T<T_1$) is, on the other hand, most 
probably due to either the continuous (static) distribution of ferromagnetic
interaction strengths (due to the presence of localised electrons) or to spin 
fluctuations in
the single ordered phase, with the fluctuation rate comparable to the nuclear
magnetic Larmor frequency\cite{nowik}. 
We note that the slow 
relaxation processes (also reported earlier, see {\it e.g.} 
ref. \cite{lynnlcmo}) may also be due to the presence of nearly-localised 
electrons. 

We next turn to the temperature dependence of the inverse magnetisation 
(shown in fig. \ref{fig:magnet} for PSMO and PSMO-F) and
notice that it deviates downward from the Curie--Weiss law, 
$H/M \propto T-T_{CW}$, when approaching $T_C$ from the paramagnetic side
(cf. ``magnetisation kink'' \cite{Salamon}). In all cases, the inverse ac 
susceptibility shows the same feature (see  fig. \ref{fig:magnet}, 
bottom panel).
This deviation starts at a point which can be recognised as the metal-insulator
transition temperature, $T_{MI}$ (within 2-3K accuracy). We emphasise that
this (and {\it not} the Curie ferro- to paramagnetic transition!) is a 
manifestation of the resistivity peak in the {\it magnetic} properties
of the system.

This downturn of $1/M(T)$ is often 
identified\cite{griffiths} as a Griffiths
singularity (denoted $T_G$). This implies that at $T_C<T<T_G$, the inhomogeneity 
of effective exchange couplings gives rise to (fluctuating) ferromagnetic
metallic clusters, resulting in the resistivity downturn at $T=T_{MI}<T_G$. 
We note 
that one does expect an 
inhomogeneity of exchanges in a system where carrier 
localisation leads to an inhomogeneous (on the atomic length scale) charge 
distribution; this probably 
accounts for the deviation of $1/M$ from the
high-temperature Curie--Weiss law, characterised by $T_{CW}^\infty>T_{CW}$.
This deviation, however, occurs at much higher temperatures, 
$T/T_C\sim 1.5-2$ \cite{deteresa}. As for the ferromagnetic cluster formation
(and hence for the possibility of the Griffiths phase),
our results as outlined above allow for this only below the temperature 
$T_2$, which in the case of PSMO-F is well below $T_{MI}$ 
(see table \ref{tab:samples}).  
%The 
%Griffiths phase approach
%may thus prove relevant for the analysis of  magnetic critical properties of
%the manganites near $T_C$, but {\it not} for understanding the origins of the
%resistivity peak at $T=T_{MI}$.

This behaviour of $1/M(T)$ is, on the other hand, easy to understand within 
the two-fluid approach.
The itinerant carriers, which would appear 
once the temperature approaches $T_{MI}$ from above, strengthen the 
double-exchange ferromagnetism\footnotemark[2].
%\cite{local}. 
This gives 
rise\footnote{We note that in  ref. \cite{Salamon}, 
the ``magnetisation kinks'' were  attributed to 
the appearance of metallic islands (cf. our $T_2$).}
to the increase of $T_{CW}$
%\cite{cfsal} 
below $T_{MI}$; we note that our plots of $1/M(T)$ 
suggest the possible
presence at $T_C<T<T_{MI}$ of an intermediate Curie--Weiss behaviour, 
$1/M \propto  T-T_{CW}^\prime$ (with $T_{CW}^\prime$ slightly larger than 
$T_C$, as expected for a Curie--Weiss temperature in a conventional magnet). 
While the abrupt
change of $T_{CW}$ at $T=T_{MI}$ would suggest the first-order nature of
the metal-insulator transition, we leave this question for future study.
%We further comment that when only a small number of carriers is released
%from the localised states into the itinerant band, they may still be
%localised on a larger length scale due to other effects (disorder),
%placing the actual $T_{MI}$ value slightly below the temperature, 
%corresponding to the downturn of $1/M$. 

\section{Conclusion}
The results presented here shed light on the nature of the electronic
states in the ferro- and paramagnetic phases of colossal magnetoresistive
manganites. In the ferromagnetic phase, the unusual strong M\"ossbauer
line broadening (observed in both PSMO and LCMO at $T>90{\rm K}$) is 
compatible with the 
presence of some localised carriers in a homogeneous metallic system.
As for the paramagnetic phase, our data for PSMO show that in the
vicinity of transition, the presence of some extended carrier states
does not necessarily lead to inhomogeneity and phase separation.

It should be stressed that our results do {\it not} imply different
physics underlying the resistivity peak in the two compounds, PSMO and LCMO.
Indeed, the peak is a generic consequence of metallic and insulating behaviour
below and above the critical temperature region respectively.
Whether the actual transition has prominent first-order characteristics
and may be accompanied by phase separation (as in the case of LCMO) 
or occurs smoothly without breaking the homogeneity of the system (PSMO)
depends on the chemical composition and doping level of the sample.
The critical behaviour of a system with multiple degrees of freedom can be
expected to vary depending on the details of the balance between different
interactions. In the case of manganites, such variation has already 
been uncovered by the neutron scattering measurements and 
magnetometry\cite{lynnlcmo}.

In summary, we found strong evidence against phase separation being
a generic mechanism 
for the resistivity peak (and hence for the CMR phenomenon). 
We suggest that the
relationship between M\"{o}ssbauer and transport data,
as well as the presence of two distinct transitions at $T_C$
and $T_{MI}$ (the latter also affecting  the magnetic properties)
can be understood within the two-fluid model.

\acknowledgments
We thank R. Berkovits, M. Golosovsky, L. Klein, E. M. Kogan,  D. Orgad, 
V. Orlyanchik,  O. A. Petrenko, and M. B. Salamon
for illuminating discussions, and members of technical staff at our
institutions for their valuable help. We acknowledge support of 
%ISF Grants \# 618/04 and \#249/05,
ISF 2004 Grant \# 618/04, ISF 2005 Grant \#249/05,
and of the Israeli Absorption Ministry.


\begin{thebibliography}{999}
\bibitem{Tokurabook} \Editor{Tokura Y.} \Book{Colossal Magnetoresistive
Oxides}
\Publ{Gordon and Breach, New York}
\Year{2000}, and references 
therein.
\bibitem{pseudogap} \Name{Saitoh T. {\it et al.}}
\REVIEW{Phys. Rev. B} {62} {2000}{1039};

\Name{Dessau D. S. \and Shen Z.-X.} in ref. \cite{Tokurabook}; 

\Name{Chuang Y.-D. {\it et al.}} \REVIEW{ Science}{292}{2001}{1509};
 
\Name{Biswas A. {\it et al.}} \REVIEW{Phys. Rev. B} {59}{1999} {5368};
 
\Name{Mitra J. {\it et al.}} \REVIEW{Phys. Rev. B} {71}{2005}{094426}. 
%note that optical [\Name{Okimoto Y. {\it
%et al}}, \REVIEW{Phys. Rev. Lett.} { 75}{1995}{109}] and transport 
%[\Name{Chun S. H.{\it et al}} \REVIEW{Physica B} {284-288} {2001}{1442}] data
%suggest that the effective carrier number decreases with $T$ increasing
%toward $T_C$.
\bibitem{okimoto}\Name{Okimoto Y. {\it
et al.}}, \REVIEW{Phys. Rev. Lett.} { 75}{1995}{109}.
\bibitem{chun}\Name{Chun S. H. {\it et al.}} \REVIEW{Physica B} 
{284-288} {2001}{1442}.
\bibitem{centralpeak} \Name{Lynn J. W. {\it et al.}} 
\REVIEW{Phys. Rev. Lett.} {76}{1996} {4046};

\Name{ Fernandez-Baca J. A. {\it et al.}} \REVIEW{Phys. Rev. Lett.} {80}{1998}{4012};

\Name{Zhang J. {\it et al.}} \REVIEW{J. Phys.: Condens. Matter} { 19} 
{2007}{315204}.
\bibitem{Nagaevreview} \Name{Nagaev E. L.} \REVIEW{Phys. Rep.} {346}
{2001}{388}.
%and references therein.
%Colossal magnetoresistance and phase
%separation in magnetic semiconductors (Imperial College Press, 2002). 
\bibitem{yamtc} \Name{Lofland S. E. {\it et al.}} \REVIEW{Phys. Rev. B} {56}
{1997}{13705}.
\bibitem{phasep} \Name{Moreo A., Yunoki S. \and  Dagotto E.}
\REVIEW{Science} { 283} {1999} {2034}; 

\Name{Uehara M. {\it et al.}}
\REVIEW{Nature} { 399}{1999}{560}.
\bibitem{Dagotto} 
\Name{Dagotto E., Hotta T. \and Moreo A.}\REVIEW{Phys. Rep.} {344}{2001}{1}.
\bibitem{Rama} \Name{Ramakrishnan T. V. {\it et al.}} 
\REVIEW{Phys. Rev. Lett.} {92}{2004}{157203}; 

\Name{Ramakrishnan T. V.} \REVIEW{J. Phys.: Condens. Matt.} 
{19}{2007}{125211}.
\bibitem{Salamon} \Name{Jaime M. {\it et al.}} \REVIEW{Phys. Rev. B}{60}{1999}
 {1028}.
\bibitem{Millis} \Name{Millis A. J.,  Mueller R.  \and Shraiman B. I.} 
\REVIEW{Phys. Rev. B}
{54}{1996}{ 5405}.
\bibitem{degennes} \Name{de Gennes P.-G.} \REVIEW{Phys. Rev.} {118}{1960}{141}. \bibitem{Millis95} {\Name{Millis A. J.,  Littlewood P. B.  \and Shraiman B. I.}
\REVIEW{Phys. Rev. Lett.} {74}{1995}{5144}.}
%\bibitem{local} Unless the electron is localised on a single lattice site, the
%localised carriers also contribute to double exchange ferromagnetism. This
%effect is however weaker than for the itinerant electrons.
\bibitem{moessgeneral} See, {\it e.g.,} 

\Name{ Chechersky V. {\it et al.}}
\REVIEW{Phys. Rev. B}
{62}{2000}{5316};

\Name{Goya F. {\it et al.}} \REVIEW{J. Appl. Phys.} {91}{2002} 
{7932}. 
\bibitem{ChecherskyFNT} \Name{Chechersky V. {\it et al.}} 
\REVIEW{Low Temp. Phys.} {23}{1997}{549}.
%{\bf 23}, 545 (1997), and
\bibitem{greeks} \Name{Assaridis E. {\it et al.}} \REVIEW{Phys. Rev. B}
{75}{2007}{224412}.
%\bibitem{muon} Other local magnetic probes,
%such as muon scattering [\Name{Heffner R. H. {\it et al}} 
%\REVIEW{Phys. Rev. Lett.} {85} {2000}
%{3285}] and nuclear magnetic resonance [\Name{Papavassiliou G. {\it et al}} 
%\REVIEW{Phys. Rev. Lett.} {84}{2000}{761}], also find several different 
%contributions in the region around $T_C$.
\bibitem{muon} \Name{Heffner R. H. {\it et al.}} \REVIEW{Phys. Rev. Lett.} {85} {2000}{3285}.
\bibitem{nmr} \Name{Papavassiliou G. {\it et al.}} 
\REVIEW{Phys. Rev. Lett.} {84}{2000}{761};

\Name{Savosta M. M. \and Nov\'{a}k P.} \REVIEW{Phys. Rev. Lett.}{87}{2001}
{137204}.
\bibitem{furnace} \Name{Shulyatev D. A. {\it et al.}} \REVIEW{J. Crystal Growth} 
{237-239}{2002}{810}.
\bibitem{iron} \Name{Ogale S. B. {\it et al.}} \REVIEW{Phys. Rev. B}  {B57}
{1998}{7841};

\Name{Li J. {\it et al.}} \REVIEW{J. Phys.: Condens. Matt.} {16}{2004}{2839}.
%\bibitem{psmof} The resistivity of PSMO (fig. \ref{fig:mit}, top) is
%indeed very close to that of PSMO-F; see also table \ref{tab:samples}. 
%\bibitem{chi} 
%We defined $T_C$ as a point where the derivative of the magnetic 
%susceptibility, 
%${\it \partial {\rm Re} \chi /\partial T}$, is minimal.
\bibitem{Schiffer} \Name{Schiffer P. {\it et al.}} \REVIEW{Phys. Rev. Lett.}
{75}{1995} {3336}. 
\bibitem{griffiths}\Name{Salamon M. B. \and Chun S. H.} \REVIEW{Phys. Rev. B}
{68}{2003}{014411}. 
%M. B. Salamon, P. Lin, and S. H. Chun, Phys. 
%Rev. Lett. {\bf 88}, 197203 (2002); 
\bibitem{McLachlan}\Name{McLachlan D. S., Blaskiewicz M. \and Newnham R. E.}
\REVIEW{J. Amer. Ceram. Soc.}{73}{1990}{2187}.
\bibitem{lynnlcmo} \Name{Adams C. P. {\it et al.}} \REVIEW{Phys. Rev. B}
{70}{2004}{134414}; 

\Name{Li W. {\it et al.}} \REVIEW{J. Phys.: Condens. Matt.} {16} {2004}{L109}.
\bibitem{moesssingle}\Name{Chechersky V. {\it et al.}} \REVIEW{Phys. Rev. B}{63}{2001}
{214401}. 
%We are aware of a sole exception 
%[\Name{Chechersky V. {\it et al}} \REVIEW{Phys. Rev. B}{63}{2001}
%{214401}], reporting emission M\"{o}ssbauer data for a single crystal of 
%${\rm La_{0.9} Ca_{0.1} Mn O_3}$, 
%which is well outside the metallic doping range.
\bibitem{nowik} \Name{Nowik I. \and Wickman H. H.} \REVIEW{Phys. Rev. Lett.}
 {17}{1966}{949}.
\bibitem{deteresa} \Name{De Teresa J. M. {\it et al.}} \REVIEW{Nature} {386}
{1997} {256}.
%\bibitem{cfsal} We note that in  ref. \cite{Salamon}, 
%the ``magnetisation kinks'' were  attributed to 
%the appearance of metallic islands (our $T_2$). 

\end{thebibliography}
\end{document}